\documentclass{article}

\usepackage{graphicx}
\usepackage{amsmath}
\usepackage[final]{gpsmdms_2022}
\usepackage{booktabs}
\usepackage{caption}
\usepackage{subcaption}
\usepackage{comment}
\usepackage{hyperref}
\usepackage{multirow}
\usepackage{lipsum}  
\usepackage{float}
\hypersetup{
    colorlinks=true,
    linkcolor=black,
    filecolor=black, 
    citecolor= black,
    urlcolor=black,
    pdftitle={Overleaf Example},
    pdfpagemode=FullScreen,
    }

\title{ Challenges in Gaussian Processes for Non Intrusive Load Monitoring}

\author{%
Aadesh Desai \quad Gautam Vashishtha \quad Zeel B Patel \quad  Nipun Batra \\\\
IIT Gandhinagar, India\\
\texttt{\{desai.aadesh,gautam.pv,patel\_zeel,nipun.batra\}@iitgn.ac.in}
}

\begin{document}
\vspace{-15pt}
\maketitle
\vspace{-25pt}
\begin{abstract}
\vspace{-10pt}
Non-intrusive load monitoring (NILM) or energy disaggregation aims to break down total household energy consumption into constituent appliances. Prior work has shown that providing an energy breakdown can help people save up to 15\% of energy. In recent years, deep neural networks (deep NNs) have made remarkable progress in the domain of NILM. In this paper, we demonstrate the performance of Gaussian Processes (GPs) for NILM. We choose GPs due to three main reasons: i) GPs inherently model uncertainty; ii) equivalence between infinite NNs and GPs; iii) by appropriately designing the kernel we can incorporate domain expertise. We explore and present the challenges of applying our GP approaches to NILM.
    
\end{abstract}

\section{Introduction}
Prior work has shown that providing an energy breakdown: per-appliance energy consumption can empower users to reduce energy consumption by up to 15\%~\cite{darby}. The most accurate way to get an energy breakdown is via ``disaggregating" the household total energy consumption measured via a smart meter, which is a single point of instrumentation per home. This line of work is often called energy disaggregation or non-intrusive load monitoring (called non-intrusive, as there is no instrumentation or intrusion in the home)~\cite{mariano}. 

The initial work in the domain involved heuristics and edge-detection-based methods~\cite{192069}. In the past decade, graphical model variants such as the factorial hidden Markov model were proposed~\cite{kelly}. However, in the recent past, deep neural networks have outperformed conventional methods. These models leverage the advances in NNs across various domains such as attention. 

In this work, we explore Gaussian Processes (GPs) for energy disaggregation. Our motivation is threefold: i) GPs inherently model uncertainty. Recent work on applying Bayesian neural networks for NILM suggests poor calibration performance~\cite{vib}; ii) given the equivalence between infinite NNs and GPs~\cite{neal}, we may expect to get comparable or better performance for GPs; iii) GPs can incorporate domain expertise via kernel design~\cite{patel}.

\section{Methods}


Given the mains (aggregate) power consumption at time $\overline{y}_t$ at time $t$, our aim is to estimate the power $\hat{y}^n_t$ for the $n^{th}$ appliance. We now introduce our GP variants. In this paper, we fit a separate GP per appliance. We summarise our models in Table~\ref{tab:kernel-table}. We also note that due to prohibitive training and testing time and memory requirement for exact GPs, we use sparse GPs~\cite{titsias2009variational} for all our GP variants. We also use automatic relevance determination (A.R.D.) for models that take a vector input.

\textbf{Point-to-Point}: In our first model, we fit a GP with the aggregate power at time $t$ as the input and the appliance power draw $\hat{y}^n_t$ as the output.


\textbf{Sequence-to-Point}: Inspired by the state-of-the-art NN model for NILM~\cite{zhang}, in our next model, the input to the GP is a window of main readings around time $t$, while the output from the GP is the appliance reading at time $t$. With such a model, we hope to capture the context of the aggregate reading. As an example, aggregate power reading of 100 Watts followed by multiple readings of 0 Watts can indicate an appliance turning off, whereas the aggregate power reading of 100 Watts followed by similar power readings indicate the appliances remaining in the same state. 

\textbf{Feature Based}: Our next model is inspired from the time series based models which incorporate advanced statistical features~\cite{ztf}. We create a latent representation of the 2$k$+1 length sequence of points that were previously used in the sequence to point model using the following features shown in Table~\ref{tab:kernel-table}. Such statistical features may provide hints to our model. For example, a high minimum value from the 2$k$+1 length sequence may indicate that multiple appliances are On during that window (higher base load). Similarly, a low value of range (difference between maximum and minimum power) may indicate that the appliance has not changed state.

\begin{table}[]
\centering
\begin{tabular}{@{}llll@{}}
\toprule
Model                & \begin{tabular}[c]{@{}l@{}}Input \\ features\end{tabular} & Kernel                & Output \\ \midrule
Point-to-point GP    &   $\overline{y}_t$                                                        & Matern (5/2)          &  $\hat{y}^n_t$      \\ \hline
Sequence-to-point GP & \begin{tabular}[c] {@{}l@{}}Sequence of 2k+1 mains readings \\($\overline{y}_{t-k:t+k}$) \end{tabular}                                                    & \begin{tabular}[c] {@{}l@{}} Matern (5/2) \\ with A.R.D. \end{tabular}  &  $\hat{y}^n_t$      \\ \hline
Feature Based GP     &  \begin{tabular}[c]{@{}l@{}}$\overline{y}_t$, max($\overline{y}_{t-k:t+k}$), min($\overline{y}_{t-k:t+k}$), \\mean($\overline{y}_{t-k:t+k}$), Kurtosis($\overline{y}_{t-k:t+k}$),\\ Difference: $\overline{y}_{t}$ - $\overline{y}_{t-1}$, range($\overline{y}_{t-k:t+k}$) \end{tabular}                                                            & \begin{tabular}[c]{@{}l@{}}Matern (5/2) \\ with A.R.D. \end{tabular} & $\hat{y}^n_t$       \\ \bottomrule
\end{tabular}
\caption{Kernels, the input features and output for each GP model}
\label{tab:kernel-table}
\end{table}

\textbf{Linear Kernel}: The central problem we face in all of our models is the incorrect prediction of the OFF state (we get an positive offset in all cases). In order to create a trainable parameter that can counter this offset, we incorporate the ``range'' feature inside a Linear Kernel in order to provide explicit hints to our model about its minimum and maximum state values. We train sequence-to-point as well feature-based model with this modification.


\section{Evaluation}

\subsection{Dataset}
We use the publicly available REDD~\cite{redd, shastri1} dataset for our evaluation. The REDD dataset consists of data from six homes over several weeks. In this work we use data from three homes similar to prior work~\cite{shastri1}. The mains data is at one second frequency, and the appliance data is at three seconds frequency. We resample both the mains and appliance data to a minute resolution, as it is a common practice in the domain~\cite{nilmtk}. The mains data consists of apparent power (vector sum of active and reactive power), whereas the appliance data consists of active power.


\subsection{Experimental Setup}

In this paper, we create an artificial aggregate instead of a true aggregate, similar to prior work~\cite{nilmtk}. The artificial aggregate is created by summing up the Refrigerator, Dishwasher and Microwave power at a given time. We also note that using the artificial aggregate simplifies the problem. 

Additionally, to test the robustness of our models to changes in the power distribution during test time, we modify the test aggregate power by adding a constant 100 Watts load. We leave the true test appliance power as it is. This experiment helps simulate the scenario when an unseen load is present in the test home.

For the sequence-to-point model, we have chosen the sequence length as 99 as used by prior work~\cite{batra}. Our GP kernel parameters are initialized from a standard normal distribution. We perform hyperparameter tuning using grid search to find optimal inducing points, learning rate and epochs for each model. The metric used to evaluate is mean absolute error (MAE)=$ \frac{1}{n}\sum_{i=1}^{n}|\hat{y}_i-y_i|$. We also report the uncertainty using the following metrics: i) CE: is calculated as the absolute difference between x and the number of samples falling within the x\% confidence interval. We report CE for a 95\% confidence interval (Hatalis et al. 2017); ii) MSLL: Mean Standardized Log Loss (MSLL) is an average of log pdf over all the test points considering the predictive distribution (Rasmussen and Williams 2005). Table \ref{tab:mae} shows the main results of our experiments. In all our approaches we use leave-one-home-out cross-validation~\cite{shastri1}. 

In the interest of space and time, we only present the results on refrigerator in this paper. Our work is fully reproducible and can be found at \href{https://github.com/Aadesh-1404/nilm\_gp}{https://github.com/Aadesh-1404/nilm\_gp}.


\begin{table}[t]
\begin{tabular}{@{}lrrrrrr@{}}
\toprule
\multirow{3}{*}{Model} & \multicolumn{6}{c}{Dataset} \\ \cmidrule(l){2-7} 
 & \multicolumn{3}{l}{Artificial Mains} & \multicolumn{3}{l}{Artificial Mains + 100} \\
 & \multicolumn{1}{l}{MAE} & \multicolumn{1}{l}{MSLL} & \multicolumn{1}{l}{CE(95\%)} & \multicolumn{1}{l}{MAE} & \multicolumn{1}{l}{MSLL} & \multicolumn{1}{l}{CE(95\%)} \\ \cmidrule(r){1-7}
Point-to-point & 18.43 & 9.41 & 0.036 & 78.89 & 38.68 & 0.033 \\
Sequence-to-point & 8.76 & 5.26 & 0.049 & 92.18 & 47.91 & 0.007 \\
Sequence-to-point + Linear Kernel & 10.46 & 5.50 & 0.047 & 88.62 & 43.44 & 0.013 \\
Statistical Features & 7.60 & 5.13 & 0.045 & 84.24 & 41.15 & 0.005 \\
Statistical Features + Linear Kernel & 7.13 & 5.03 & 0.046 & 73.48 & 33.52 & 0.030 \\ \bottomrule
\end{tabular}
\caption{Metrics for different GP methods.}
\label{tab:mae}
\end{table}
\subsection{Results and Analysis}
Our main result is shown in Table~\ref{tab:mae}. We first discuss the results on artificial aggregate followed by artificial aggregate + test bias dataset. Overall, we find our statistical features model with the linear kernel on the range performs the best. 
\begin{figure}[h]
  \centering
  \begin{subfigure}[b]{0.3\textwidth}
    \includegraphics[width=\linewidth]{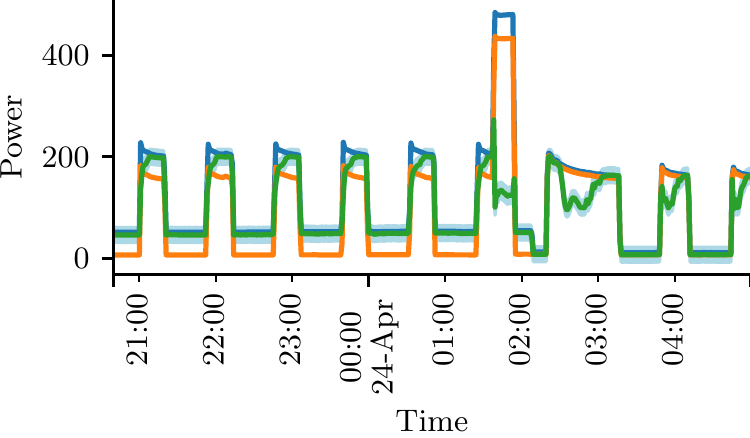}
    \caption{ }
  \end{subfigure}
   \begin{subfigure}[b]{0.3\textwidth}
    \includegraphics[width=\linewidth]{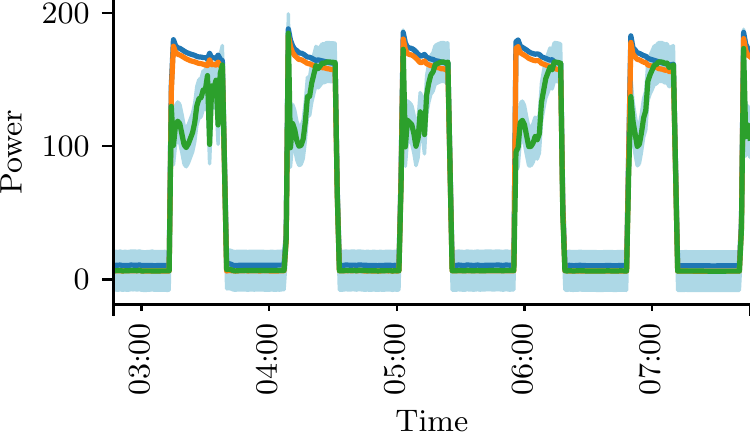}
    \caption{ }
  \end{subfigure}
   \begin{subfigure}[b]{0.3\textwidth}
    \includegraphics[width=\linewidth]{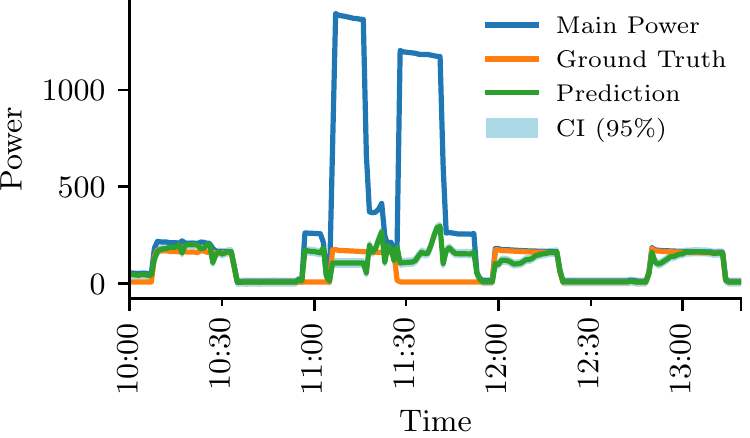}
    \caption{}
  \end{subfigure}
    \vspace{-5pt}

  \caption{Predictions from the Point-to-point model}
  \label{fig:fig1}
\end{figure}

\subsubsection{Artificial aggregate}

\begin{figure}[H]
  \centering
  \begin{subfigure}[b]{0.4\textwidth}
    \includegraphics[width=\linewidth]{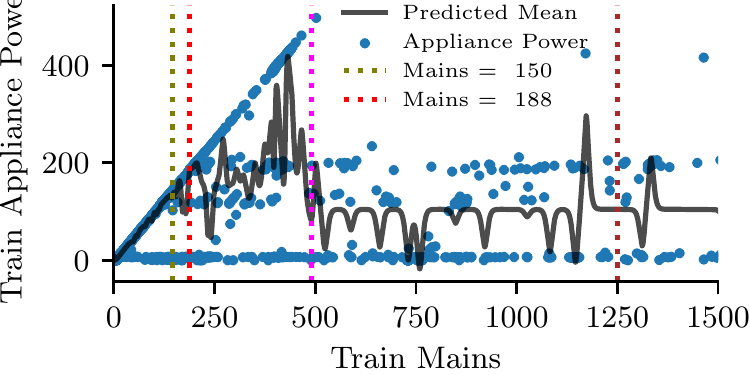}
    \caption{}
  \end{subfigure}
   \begin{subfigure}[b]{0.3\textwidth}
    \includegraphics[width=\linewidth]{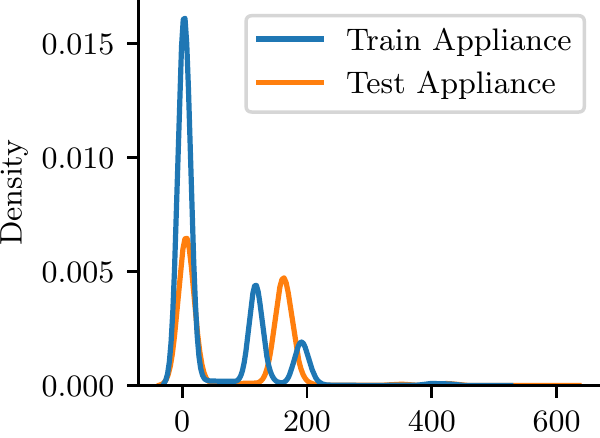}
    \caption{ }
  \end{subfigure}
   \begin{subfigure}[b]{0.2\textwidth}
    \includegraphics[width=\linewidth]{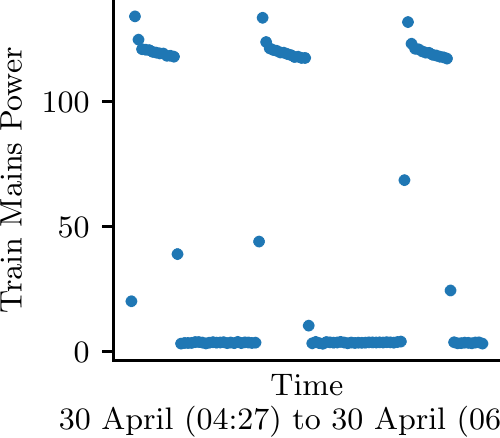}
    \caption{}
  \end{subfigure}
  \vspace{-5pt}
  \caption{Point-to-point Analysis: (a) Prediction of the GP with the training data (b) Density plot for appliance power distribution (c) scatter plot of train mains power}
  \label{fig:fig2}
\end{figure}

\textbf{Point-to-Point:}
Figure~\ref{fig:fig1} shows the predictions for our point-to-point GP model for three different windows of time. In Figure~\ref{fig:fig1}(a), we can note that the ground truth fridge power is 0 Watts when the fridge is OFF (e.g. 22:30 to 23:00 hours). However, our model predicts the power as 40 Watts. This can be explained using ~\ref{fig:fig2}(a), which shows the training appliance vs. aggregate data and the GP fit. 3 patterns can be observed in this figure: 1) the y=x line denotes the case when the only refrigerator is turned ON, 
2) y=K (around 200 Watts) line denotes the case when multiple devices are turned ON along with the refrigerator and 
3) y=0 line denotes when the refrigerator power is 0, but other appliances are ON. 
Thus, we can explain our model's non-zero (approximately 40 Watts) prediction using these 3 states. The constant fluctuation between these three states drives the prediction of the model to an intermediate value in this range predicting to 0 Watts. Figure~\ref{fig:fig1} (b) depicts the case when the refrigerator is the only active appliance (and thus, mains power is equal to refrigerator power). We observe a certain amount of dip at the rising edge of every cycle. We can explain this from the scatter plot in Figure~\ref{fig:fig2} (c), which illustrates the inadequate number of data points in the rising edge, faltering the model's predictive abilities in that region.

At first, reaching the ON state (188 W), the value falls to a lower value corresponding to mains (188 W) as depicted in Figure~\ref{fig:fig2} (a), then again rises to the value corresponding to mains (150 W). Figure~\ref{fig:fig1} (c) illustrates the prediction of the model at higher values of  power. It can be observed that there is vast fluctuation in the predictions in this region which can be reasoned by the density plots shown in Figure~\ref{fig:fig2} (b). The test readings have a higher density around 200 Watts compared to the train readings and thus the GP model is unable to make accurate predictions in this power range.

\begin{figure}[h]
  \centering
  \begin{subfigure}[b]{0.3\textwidth}
    \includegraphics[width=\linewidth]{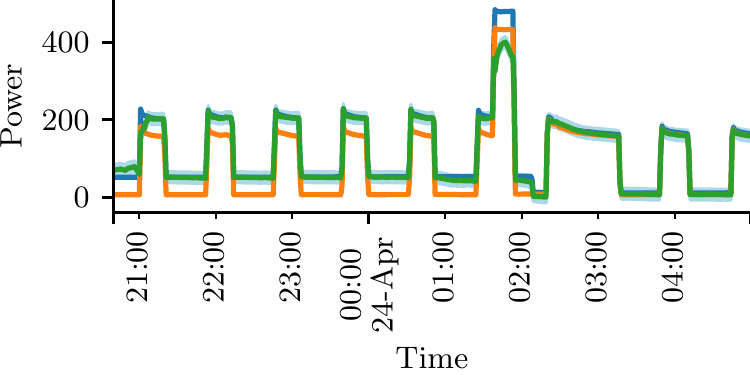}
    \caption{ }
  \end{subfigure}
   \begin{subfigure}[b]{0.3\textwidth}
    \includegraphics[width=\linewidth]{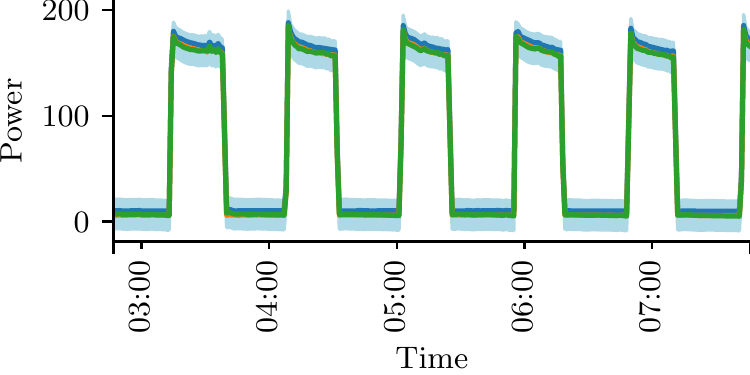}
    \caption{ }
  \end{subfigure}
   \begin{subfigure}[b]{0.3\textwidth}
    \includegraphics[width=\linewidth]{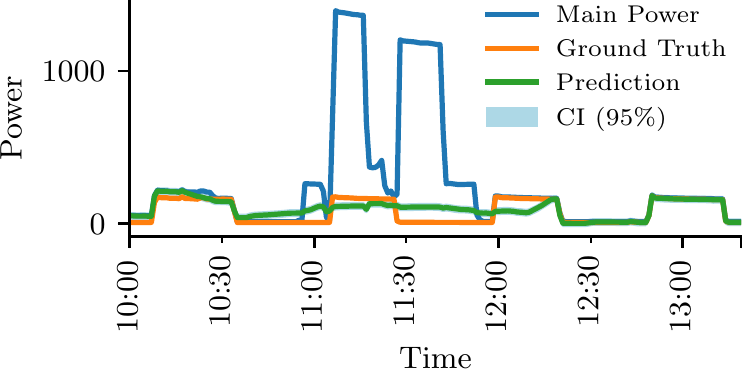}
    \caption{ }
  \end{subfigure}
  \vspace{-5pt}
  \caption{Predictions for the Sequence-to-sequence model}
  \label{fig:fig3}
\end{figure}

\textbf{Sequence-to-Point:} This model performs significantly better than the point-to-point approach. One of the earlier mentioned challenges of the point-to-point model was predicting the ON state values correctly when the fridge was the only ON appliance. Figure~\ref{fig:fig3} (b) shows that the Sequence-to-point model performs significantly better than the point-to-point model (Figure~\ref{fig:fig1 (b))}. 
Also, from \ref{fig:fig3} (c), we can note that when observing sudden/unseen changes in the training mains, the GP prediction are smooth, which was not the case in Point-to-point Figure~\ref{fig:fig1} (b). However, as seen from \ref{fig:fig3} (a), though predictions for defrost stage improve, the model still fails to accurately predict the OFF stage.

\begin{figure}[H]
  \centering
    \begin{subfigure}[b]{0.2\textwidth}
    \includegraphics[width=\linewidth]{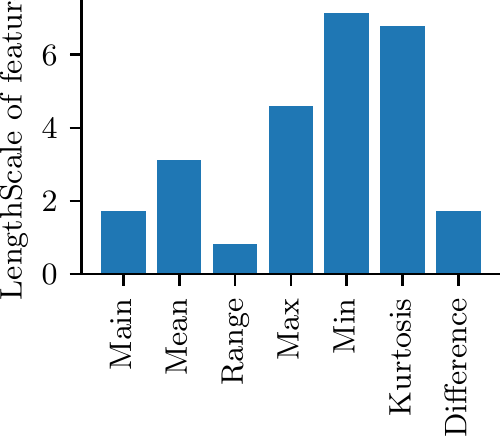}
    \caption{ }
  \end{subfigure}
    \begin{subfigure}[b]{0.4\textwidth}
    \includegraphics[width=\linewidth]{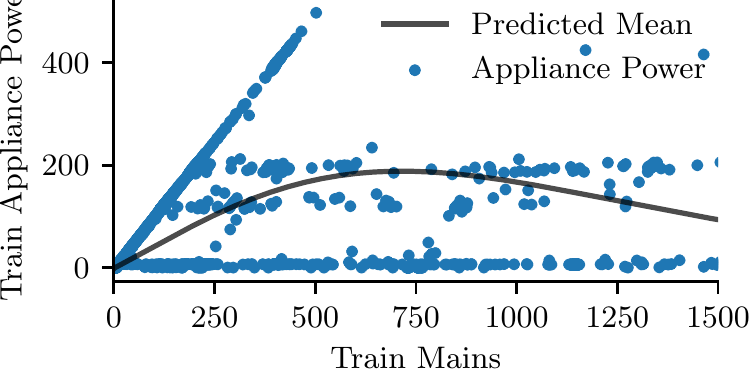}
    \caption{ }
  \end{subfigure}
  \begin{subfigure}[b]{0.3\textwidth}
    \includegraphics[width=\linewidth]{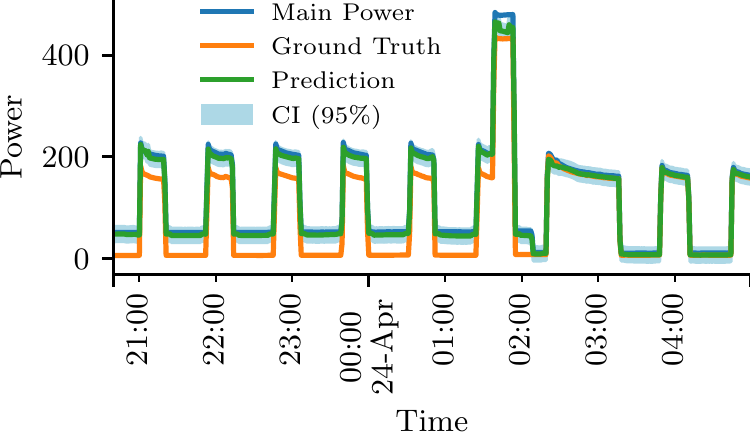}
    \caption{ }
  \end{subfigure}
  \vspace{-5pt}

  \caption{Feature based GP: (a) A.R.D plot, (b) Mean Prediction (c) Predictions showing failure }
  \label{fig:fig4}
\end{figure}

\textbf{Features to Point:}  We can observe from \ref{fig:fig4}(b) that this model performs significantly better than point-to-point GP and also beats sequence-to-point GP. \ref{fig:fig4}(c) also illustrates that we are able to predict the defrost state better using the model. From, \ref{fig:fig4}(a), we conclude Range along with difference  as the important features from the ARD plot. However, we can also see that this model still fails in correctly predicting the OFF state \ref{fig:fig4}(c).



\subsubsection{Artificial aggregate + Test Bias}
\begin{figure}[H]
  \centering

   \begin{subfigure}[b]{0.23\textwidth}
    \includegraphics[width=\linewidth]{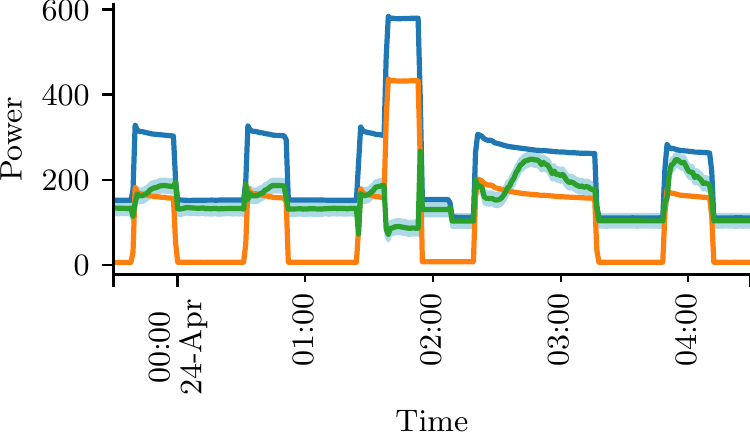}
    \caption{ }
  \end{subfigure}
  \begin{subfigure}[b]{0.23\textwidth}
    \includegraphics[width=\linewidth]{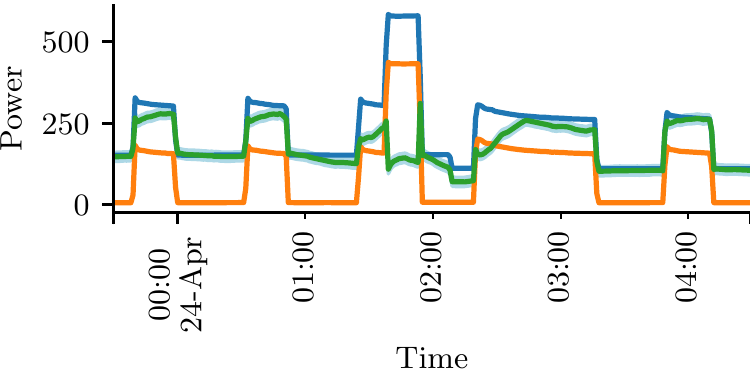}
    \caption{ }
  \end{subfigure}
  \begin{subfigure}[b]{0.23\textwidth}
    \includegraphics[width=\linewidth]{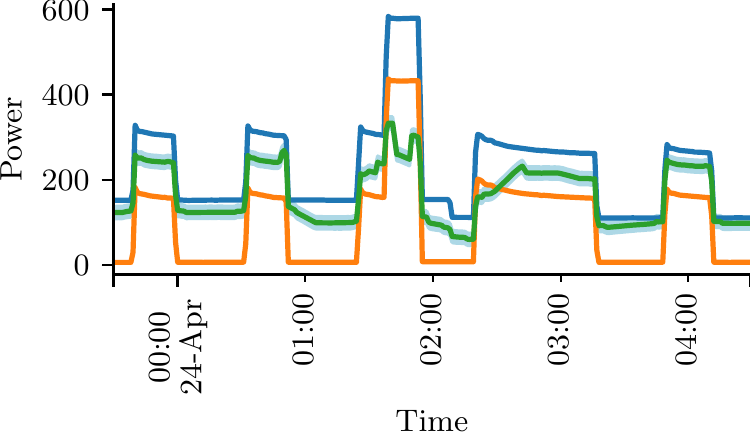}
    \caption{ }
  \end{subfigure}
     \begin{subfigure}[b]{0.23\textwidth}
    \includegraphics[width=\linewidth]{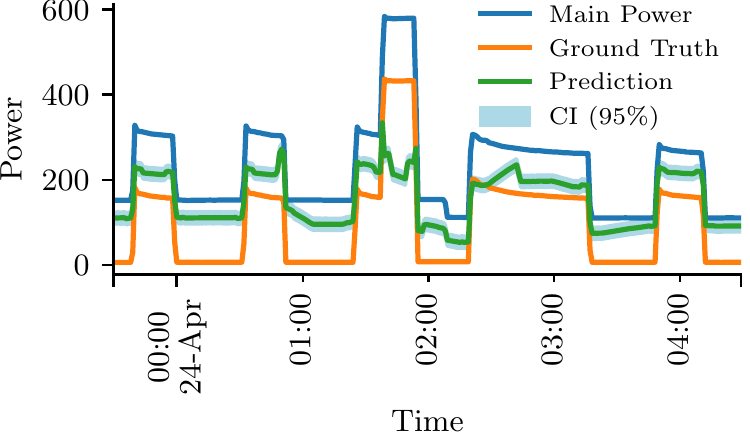}
    \caption{}
  \end{subfigure}
     \vspace{-5pt}

  \caption{GP Output on adding bias: (a) Point-to-point (b) Sequence-to-point (c) Feature based (d) Feature based + Linear Kernel}
  \label{fig:fig5}
\end{figure}

\textbf{Point-to-point} - Fig~\ref{fig:fig5} (a) illustrated on adding a bias, the results fail significantly, as the mains is shifted it simply gives predictions for per point shift as per the predicted mean in Fig~\ref{fig:fig5} (a). 

\textbf{Sequence-to-point} - this model performs worse than Point-to-point. This can be explained by the covariance matrix being formed from 99-time steps which all add a uniform bias, making the model make incorrect predictions with more certainty and thus the inferior nature of the prediction. To mitigate the bias, we create a learnable parameter by adding the range (max-min) of the mains feature and incorporating it in a linear kernel. This strategy improved predictions.

\textbf{Features to point:} This model fails as expected, but the results are better than Sequence-to-point, this is because the bias is not uniform across all the features and the distributed bias gives better predictions. On using range with linear kernel it is able to compensate for the bias (Fig~\ref{fig:fig5}(d)).

 We quantify uncertainty in prediction by plotting the reliability curves (in the Appendix) and using the expected calibration error~\cite{vib}. The reliability curve (calibration curve) can help visually understand the model’s calibration while the expected calibration error (ECE) quantitatively measures the model calibration. A lower ECE value indicates a better calibration. We have computed ECE values for our models in Table \ref{tab:ece}.

\begin{table}[H]
\begin{tabular}{@{}lrllrll@{}}
\toprule
\multirow{3}{*}{Model} & \multicolumn{6}{c}{Dataset} \\ \cmidrule(l){2-7} 
 & \multicolumn{3}{l}{\multirow{2}{*}{Artificial Mains}} & \multicolumn{3}{l}{\multirow{2}{*}{Artificial Mains + 100}} \\
 & \multicolumn{3}{l}{} & \multicolumn{3}{l}{} \\ \cmidrule(r){1-1}
Point-to-point & \multicolumn{3}{r}{0.196} & \multicolumn{3}{r}{0.483} \\
Sequence-to-point & \multicolumn{3}{r}{0.229} & \multicolumn{3}{r}{0.496} \\
Sequence-to-point + Range (Linear Kernel) & \multicolumn{3}{r}{0.085} & \multicolumn{3}{r}{0.493} \\
Statistical Features & \multicolumn{3}{r}{0.269} & \multicolumn{3}{r}{0.493} \\
Statistical Features + Range (Linear Kernel) & \multicolumn{3}{r}{0.279} & \multicolumn{3}{r}{0.486} \\ \bottomrule
\end{tabular}
\caption{Expected Calibration Error}
\label{tab:ece}
\end{table}

\section{Conclusion \& Future Work}

In this work, we investigate GPs for energy NILM. Energy disaggregation is a complex multi-trend dataset which is difficult to predict with point-to-point approaches. The sequence-to-point model performed better due to access to a larger bin of information but lacked in terms of obscure latent representation of the feature space. Explicit featurization in the feature-to-point model performed the best due to access to concise information and a semantically rich latent space. We are optimistic that significant progress lies ahead in this area of study. Advanced featurization and explicit kernel designing can be useful for appropriately disaggregating power for all the appliances. Further, we can explore Deep GP models, Deep Kernel Learning and MultiOutput GPs for NILM.

\bibliographystyle{unsrt}
\bibliography{main}

\newpage
\appendix

\section{Appendix}

\subsection{Calibration} 


\begin{figure}[h]
  \centering
    \begin{subfigure}[b]{0.3\textwidth}
    \includegraphics[width=\linewidth]{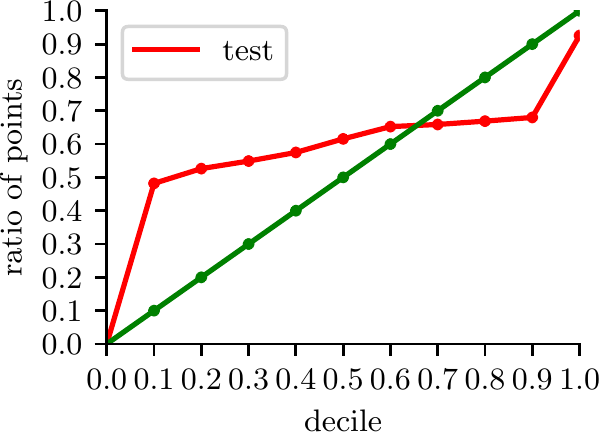}
    \caption{ }
  \end{subfigure}
    \begin{subfigure}[b]{0.3\textwidth}
    \includegraphics[width=\linewidth]{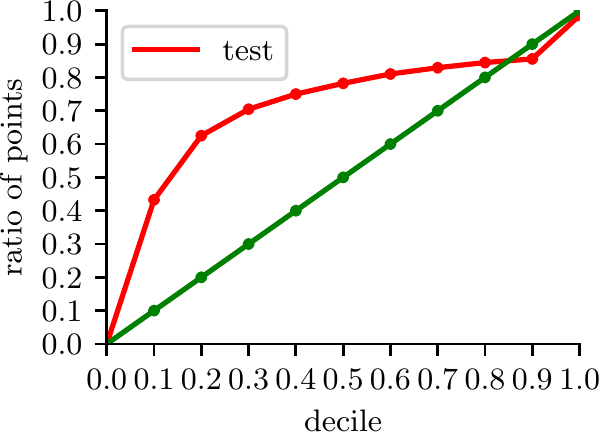}
    \caption{ }
  \end{subfigure}
  \begin{subfigure}[b]{0.3\textwidth}
    \includegraphics[width=\linewidth]{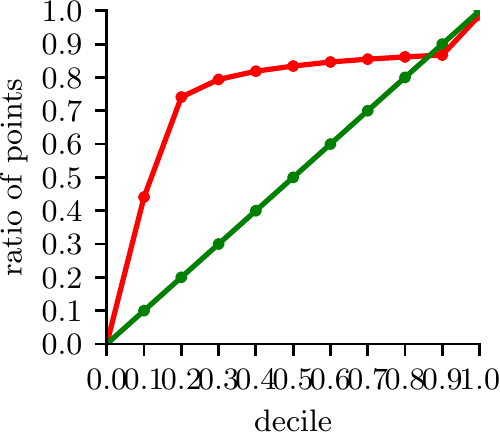}
    \caption{ }
  \end{subfigure}
  \vspace{-5pt}

  \caption{Calibration Plots: (a) Point-to-point GP, (b) Sequence-to-point GP (c) Feature Based GP }
  \label{fig:fig6}
\end{figure}

\begin{figure}[h]
  \centering
    \begin{subfigure}[b]{0.3\textwidth}
    \includegraphics[width=\linewidth]{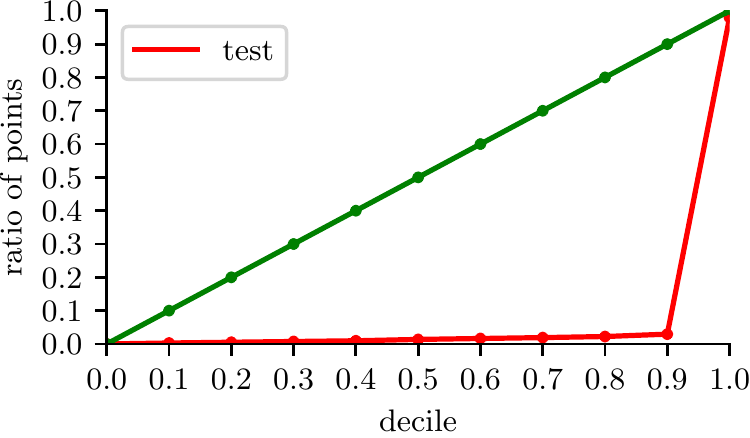}
    \caption{ }
  \end{subfigure}
    \begin{subfigure}[b]{0.3\textwidth}
    \includegraphics[width=\linewidth]{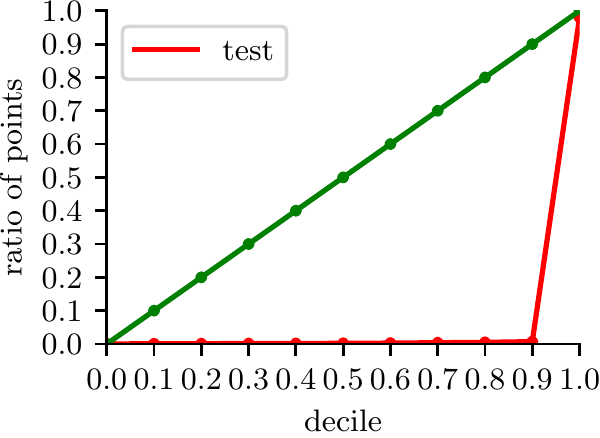}
    \caption{ }
  \end{subfigure}
  \begin{subfigure}[b]{0.3\textwidth}
    \includegraphics[width=\linewidth]{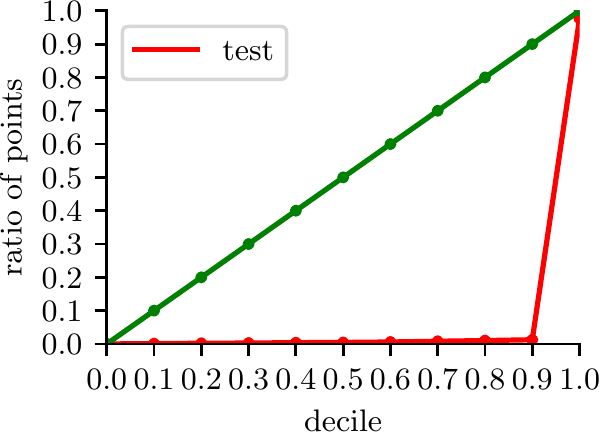}
    \caption{ }
  \end{subfigure}
  \vspace{-5pt}

  \caption{Calibration Plots Added Bias: (a) Point-to-point GP, (b) Sequence-to-point GP (c) Feature Based GP }
  \label{fig:fig7}
\end{figure}

\begin{figure}[h]
  \centering
    \begin{subfigure}[b]{0.3\textwidth}
    \includegraphics[width=\linewidth]{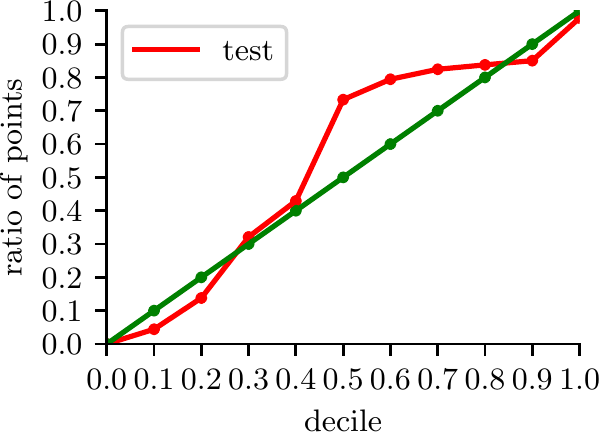}
    \caption{ }
  \end{subfigure}
    \begin{subfigure}[b]{0.3\textwidth}
    \includegraphics[width=\linewidth]{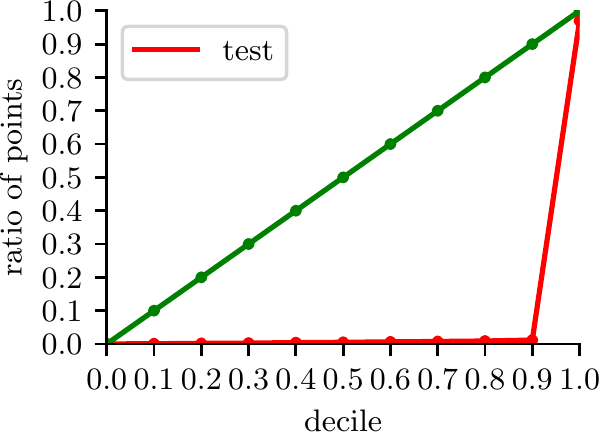}
    \caption{ }
  \end{subfigure}
  \begin{subfigure}[b]{0.3\textwidth}
    \includegraphics[width=\linewidth]{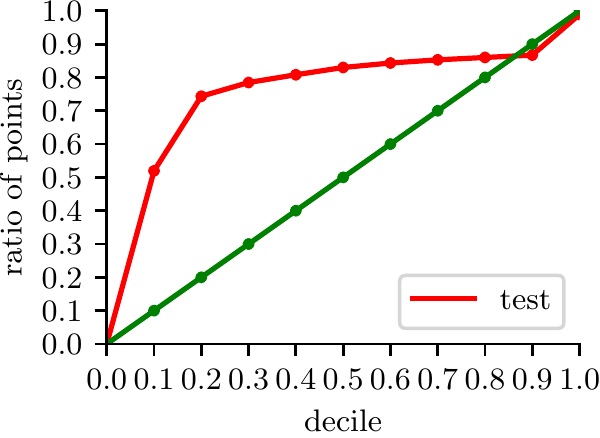}
    \caption{ }
  \end{subfigure}
  \begin{subfigure}[b]{0.3\textwidth}
    \includegraphics[width=\linewidth]{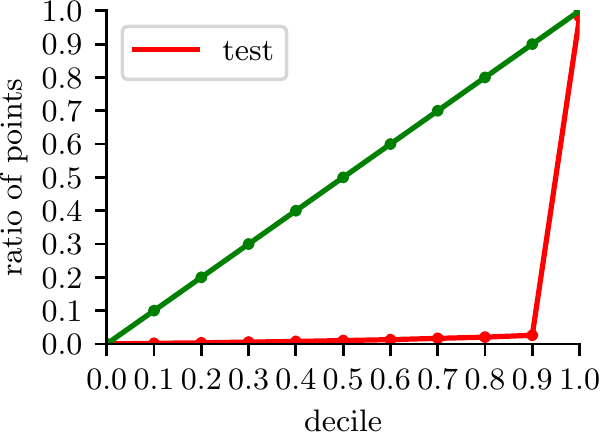}
    \caption{Statistical Features}
  \end{subfigure}
  \vspace{-5pt}

  \caption{Calibration Plots Linear: (a) Seq-to-point + Linear GP, (b) Bias: Seq-to-point + Linear GP, (c) Feature Based + Linear GP, (d) Bias: Feature Based + Linear GP }
  \label{fig:fig6}
\end{figure}


\end{document}